\documentclass[11pt]{article}
\usepackage{amsthm}
\usepackage{amsmath}
\usepackage{amssymb}
\usepackage{epsfig}

\newtheorem{myth}{\bf Theorem}
\newtheorem{myprop}[myth]{\bf Proposition}

\newtheorem{myle}[myth]{\bf Lemma}

\newtheorem{myalg}{\bf Algorithm}

\newcommand{\bc}{\boldsymbol c}
\newcommand{\bx}{\boldsymbol x}
\newcommand{\bp}{\boldsymbol p}
\newcommand{\bJ}{\boldsymbol J}
\newcommand{\bA}{\boldsymbol A}
\newcommand{\one}{\mathbf{1}}

\newcommand{\Mo}{\mathcal{M}_{-i}}

\title{\bf Information Flow in One-Dimensional Vehicular Ad~Hoc Networks}
\author{{\large Chi Wan Sung$^\dagger$, Kenneth W. Shum$^\dagger$, Wing Ho Yuen$^\ddagger$} \\
\\
\small $^\dagger$ Department of Electronic Engineering \\
\small City University of Hong Kong \\
\small $^\ddagger$ Department of Computer Science \\
\small Columbia University, New York \\
\\
{\small \tt albert.sung@cityu.edu.hk}, {\small \tt kshum2008@gmail.com}, {\small \tt andyyuen@cs.columbia.edu}}
\date{}
\begin{document}

\maketitle


\begin{abstract}
We consider content distribution in vehicular ad hoc networks. We assume that a file is encoded using fountain code, and the encoded message is cached at infostations. Vehicles are allowed to download data packets
from infostations, which are placed along a highway. In addition, two vehicles can exchange packets with each other when they are in proximity. As long as a vehicle has received enough packets from infostations or from other vehicles, the original file can be recovered. In this work, we show that
system throughput increases linearly with number of users, meaning
that the system exhibits linear scalability. Furthermore, we analyze the effect of mobility
on system throughput by considering both discrete and continuous
velocity distributions for the vehicles. In both cases, system
throughput is shown to decrease when the average speed of all vehicles
increases. In other words, higher overall mobility reduces system throughput.
\end{abstract}


\newpage

\section{Introduction}

Peer-to-peer file sharing has been a new paradigm for content
distribution over the Internet for many years. It differs from
the traditional client-server architecture in that after a client
starts downloading a file from one or more servers, it itself
becomes a server, able to serve other users' requests. In
other words, each participating node plays the role of both server
and client at the same time. There is no single bottleneck in the
system; the capacity grows when more nodes participate, resulting in
endless scalability. The best example is the highly popular
BitTorrent system~\cite{cohen}. It splits the sharing file into
small blocks. Users download missing blocks from their peers. Once
downloaded, those blocks become available to their peers.


Although many peer-to-peer file sharing systems were developed for
wired network, they may not be suitable for all kinds of wireless
systems. For example, in wireless ad hoc networks, a packet
typically has to traverse multiple hops from source to destination.
It was shown in \cite{gupta} that per-node
throughput changes at a rate $O(1/\sqrt{n \ln n})$, which drops to
zero for large $n$. Consequently, the multi-hop strategy is
intrinsically unscalable, no matter what protocols are used in the
network layer and above. To maintain scalability, a two-hop relay strategy
was considered in~\cite{gross}. It was shown that with node mobility,
per-node throughput becomes $O(1)$. As a result, system capacity scales linearly
with the number of nodes. This drastic difference motivates the design of many
mobility-assisted data transfer protocols. Some examples are
presented in~\cite{papa2,yuen}.

In this work, we focus on vehicular ad hoc networks (VANET), which consists of cars, trucks,
motorcycles, and all sorts of vehicles on the road.
A major characteristic of VANET is its highly dynamic topology.
Nodes are intermittently connected when they encounter one another
on the road. If traffic density is low, the proportion of time that
a node is connected to another node may be small, which may result
in large delay. On the other hand, the instantaneous transmission
rate can be very high, especially if transmission proceeds only when
two nodes are close to each other. Due to its nature, VANET is
particularly suitable for delay tolerant applications with large
bandwidth requirement. An example is that a content provider allows
its subscribers to download movies, music, or news from an
infostation at the roadside when they pass by, and to exchange
contents among themselves when they encounter one another on the
road. A user can simply run an application program in the background,
without the aware of download schedule. To facilitate the
development of these applications, a content distribution protocol
is needed. A BitTorrent-like protocol called CarTorrent was proposed
in~\cite{nandan}. Two other protocols were designed in~\cite{ahmed,lee} based on the
idea of network coding~\cite{raymond,li}. In this work, we adopt the
fountain code approach~\cite{mackey05}. Encoding is performed at infostations but not
at vehicles. This method can reduce processing time at vehicles, and reduce decoding
complexity if a suitable fountain code is used.

The contributions of this work are these: First, the application scenario is modeled, which reveals the relationship between coding, delay, and throughput. Second, exact formulae for throughput are derived, from which insights
on how mobility affects throughput can be gained. Our approach is similar to that in \cite{yuen2}, but with some major differences in modeling.

\section{Content Distribution for Vehicular Network}

We consider a one-dimensional vehicular network, which models the scenario where many cars are running on a highway. Suppose that a portion of car users subscribes to a content distribution network. They are interested in downloading a common file from the content provider. The file is split into $K$ smaller blocks $W_1, W_2, \ldots, W_K$, each of which consists of $L$ bits. These messages blocks are cached in infostations~\cite{frenkiel}, placed along the highway. When a car comes close to an infostation, it can download message blocks from it. Besides, a car can exchange message blocks with another car in proximity. We refer a car or an infostation as a {\em node} and say that a {\em node encounter} occurs when two nodes are approaching to within a transmit range $r$ from each other. Data exchange between the two nodes then begins. The amount of data exchange depends on the transmission bit rate $R_b$ and the connection time. We assume that non-adaptive radio is used so that $R_b$ is constant throughout the encounter period. We also assume channel coding is used so that the probability of decoding error is negligible.

We adopt a fountain code approach~\cite{mackey05} for file distribution at infostations. When a car is within the transmit range of an infostation, the infostation generates and transmits some encoded messages to the car. Each encoded message is obtained by linearly combining the original message blocks:
\begin{equation}
\sum_{k=1}^K c_k W_k, \label{encoded_message}
\end{equation}
where each $c_k$ is either 0 or 1, and the addition is performed over $\mathbb{F}_2$. The vector $\bc = (c_1, c_2, \ldots, c_K)$ is called the {\em encoding vector}, which is generated randomly. There are various ways to generate it. One simple way is to pick a vector uniformly at random over $\mathbb{F}_2^K$. Another way is to generate it according to the robust soliton distribution in LT codes~\cite{luby}. Each packet consists of an encoded message as in~\eqref{encoded_message} and the corresponding encoding vector.

The protocol that we propose for packet exchange follows a two-hop strategy. When two cars are within the transmit range of each other, they will exchange those packets that are directly downloaded from infostations. Those packets that are received from other cars will not be forwarded again. In other words, each packet is transmitted in at most two hops: from an infostation to a car, and from that car to another car.

A vehicle can recover the original file if the encoding vectors in the received packets span the vector space~$\mathbb{F}_2^K$, which happens when $K$ linearly independent encoding vectors have been received. Indeed, if $\bc_1$, $\bc_2, \ldots, \bc_K$ are encoding vectors that are linearly independent, the file can be decoded by inverting the $K\times K$ matrix whose $i$th row is $\bc_i$ for $i=1,2,\ldots, K$.

\section{Throughput Analysis}


We assume that cars arrived at the highway follow a Poisson process with rate~$\lambda$. Each of them travels in the highway at constant velocity. Those coming from the left has positive velocity and are collectively called the {\em forward traffic}. Those coming from the right has negative velocity and are called the {\em reverse traffic}.


In the highway, two nodes are connected if their distance is less than or equal to the transmit range, denoted by $r$.
The connection time between two cars, $T_c$, depends on their relative speed and is given by
\begin{equation}
T_c = \frac{r}{|v - v'|}.
\end{equation}
Note that the difference of velocity $v-v'$ may be negative, and the sign
depends on their directions. The maximum number of packets that can be
exchanged during an encounter is $R_p T_c$, where $R_p$ is the transmission rate in packets per second and is equal to  $R_b$ divided by the packet size in bits. Likewise, a car of velocity $v$ can download $R_p r/|v|$ packets from an infostation in one encounter.

We assume that there is an infostation at every entrance of the highway. When a vehicle enters the highway system, it collects some encoded message blocks from the infostation. As cars usually enter the highway at low speed, they should have picked up enough packets to be exchanged during any future encounter.  When two nodes of velocities $v$ and $v'$ meet each other, we assume that the number of packets transmitted in each direction is $R_p r/(2|v-v'|)$.

Since the velocity of each node is assumed constant, two nodes meet each other at most once as they travel along the highway. We can therefore guarantee that any newly received packet by a car is statistically independent of the packets already stored in its buffer. Consequently, the packets received by a vehicle are all statistically independently.

The number of packets that must be received before $K$ linearly independent encoding vectors are obtained depends on the probability distribution of encoding vectors. Based on the assumption that the received encoding vectors are statistically independent, the following results apply: If the distribution of encoding vectors is uniform, the original file can be decoded with probability $1-\epsilon$, for some small constant $\epsilon$, after $K+\log_2(1/\epsilon)$ packets are received. If we use LT code with robust soliton distribution, the number of packets needed is $K+2S\log_2(S/\epsilon)$, where $S=c \sqrt{K}\log_e(K/\delta)$ and $c$ is a parameter of order~1~\cite{mackey05}. Given the probability of decoding failure $\epsilon$, the downloading time is obtained by
dividing the required number of packets by the packet rate. Our objective is to estimate the average downloading time of the file in VANET by analyzing the packet rate. In the sequel, we will call it {\em throughput}. We will first consider the case where the velocity distribution is discrete, and then extend the results to the continuous case at the end of this section.

\subsection{Discrete Velocity Distribution}

Suppose that the velocity $V$ of
a vehicle can take on values from a finite set, $\{v_1, v_2, \ldots,
v_M\}$, with probability $p_1, p_2, \ldots, p_M$ respectively, where
$\sum_{m=1}^M p_m = 1$. Denote the set $\{1, 2, \ldots, M\}$ by
${\cal M}$. This model is applicable to the scenario where the
traffic is heavy and nodes using different lanes are of different
speeds. A node, when entering the highway, can choose a suitable
lane.

We consider a specific node, called the {\em
observer node}, or simply the {\em observer}, traveling between a segment of highway between two consecutive infostations \texttt{A} and~\texttt{B}. We will analysis the throughput of the observer in this segment of the highway.

Suppose that the observer belongs to class $i$ for some $i\in\mathcal{M}$, and moves at speed $v_i$ in the forward direction from \texttt{A} to~\texttt{B}. Assume that the length of this segment of the highway is~$d$. The traveling time of the observer in this segment is given by
$t_i=d/v_i$. We denote $N_i$ as the number of {\em node encounters}
for the observer when traveling in this segment of highway.
Furthermore, for $k=1,2,\ldots, N_i$, we denote $B_i(k)$ as the
number of packets received from the $k$-th encounter. Assuming that the observer does not encounter two other nodes at the same time, the total number of packets received by the observer in this highway segment is
\begin{equation}
B_i = \frac{Rr}{v_iP} + \sum_{k=1}^{N_i} B_i(k).
\end{equation}
The first term corresponds to the packets directly downloaded from infostation \texttt{A} and the second the total number of packets from other vehicles.



In order to find the expected value of $B_i$, we split the
Poisson arrival process into $M$ independent Poisson streams with
rate $p_m \lambda$, where $m=1, 2, \ldots, M$. Let $\tilde{N}_{i,m}$ be the
number of encounters of the observer with nodes in class~$m$, so that
\begin{equation}
N_i = \tilde{N}_{i,1} + \tilde{N}_{i,2} + \ldots + \tilde{N}_{i,M}.
\end{equation}
The following lemma gives the expected value of $\tilde{N}_{i,m}$.

\begin{myle} \label{le:N_m}
$\tilde{N}_{i,m}$ is Poisson distributed with mean
\begin{equation}
E[\tilde{N}_{i,m}] = \lambda p_m |t_m - t_i|,
\end{equation}
where $t_m = d/ v_m$.
\end{myle}

\begin{proof}
Without loss of generality, suppose the observer enters the highway
segment at time 0 and departs at time $t_i$. We consider its encounter
with forward traffic and reverse traffic separately.

For forward traffic, consider a node of
velocity $v_m > 0$, which enters the highway segment at time $t$ and
departs at time $t + t_m$. Suppose the speed of the node is
lower than that of the observer, that is, $t_m > t_i$. It will
encounter the observer if and only if it enters the highway before
the observer does (i.e., $t<0$) and it departs the highway after the
observer does (i.e., $t + t_m > t_i$). In other words, an encounter
occurs if and only if $-(t_m - t_i) < t < 0$. Since the arrival
process is Poisson with rate $\lambda p_m$, the number of encounters
is Poisson distributed with mean $\lambda p_m (t_m - t_i)$. Next
suppose $t_m < t_i$. An encounter occurs if and only if the node
enters after time 0 (i.e., $t>0$) and it departs before $t_i$ (i.e.,
$t + t_m < t_i$). Again the number of encounters is Poisson
distributed with mean $\lambda p_m |t_m - t_i|$.

For reverse traffic, consider a node of velocity $v_m < 0$.
If it enters the highway before time 0 (i.e., $t < 0$), it will
encounter the observer if $t+ |t_m| > 0$. If it enters the highway
after time 0 (i.e., $t > 0$), it will encounter the observer if it
enters before $t_i$ (i.e., $t < t_i$). Combining the two cases, we
can see that an encounter occurs if $-|t_m| < t < t_i$. Hence, the
number of encounters is Poisson distributed with mean also equal to
$\lambda p_m |t_m - t_i|$.
\end{proof}

Let $\Mo$ be the set ${\cal M} \setminus \{i\}$. We next obtain an
expression for the mean of $B_i$.

\begin{myle}
\begin{equation}
E[B_i] = \frac{R_p r t_i}{d}\Big[ 1  + \frac{\lambda}{2} \sum_{m \in \Mo} p_m
|t_m| \Big].
\end{equation}
\end{myle}

\begin{proof}
The observer will only encounter a node in class $m$ for $m
\neq i$. When the observer meets another node of velocity $v_m$, $v_m\neq v_i$, the number of packets received is equal to $R_p r/(2|v_m - v_i|)$. We sum over all $m\in \Mo$ and obtain
\begin{equation}
B_i  = \frac{R_p r}{v_i} + \sum_{m \in \Mo}  \frac{\tilde{N}_{i,m} R_p r}{2|v_m - v_i|}.
\end{equation}
Taking expectation and using Lemma~\ref{le:N_m}, we have
\begin{align}
E[B_i]&= \frac{ R_p r}{v_i} + \sum_{m \in \Mo}  \frac{E[\tilde{N}_{i,m}] R_p r}{2|v_m - v_i|} \\
& = R_p r \Big[\frac{1}{v_i} +  \sum_{m \in \Mo} \frac{\lambda p_m |t_m - t_i|}{2 |v_m - v_i|} \Big]\\
& = R_p r \Big[\frac{t_i}{d} +  \sum_{m \in \Mo} \frac{\lambda p_m t_i |t_m|}{2d} \Big].
\end{align}
\end{proof}

Define $C_i= B_i/ t_i$ as the average throughput of the observer during its
traveling time on the highway segment. Then we have

\begin{equation} \label{eq:main1}
E[C_i] =  \frac{R_p r}{d}\Big[ 1  + \frac{\lambda}{2} \sum_{m \in \Mo} p_m
|t_m| \Big].
\end{equation}

Consider a particular time instant $t$. A car of velocity $v_m$ will
be on this highway segment if it enters this segment within the interval $[t-|t_m|, t]$. Therefore, the number of cars of velocity $v_m$ that are on the highway is Poisson distributed with mean equal to $\lambda p_m |t_m|$.  The above equation can be rewritten in terms of car density as follows:

\begin{myth} Let $\rho_m \triangleq \lambda p_m |t_m|/d = \lambda p_m/|v_m|$ be the density of cars of velocity $v_m$. Then
\begin{equation}
E[C_i] = R_p r \left( \frac{1}{d} + \frac{1}{2}\sum_{m\in\Mo} \rho_m \right). \label{eq:discrete1}
\end{equation} \label{thm:discrete1}
\end{myth}

The first term within the parenthesis in Theorem~\ref{thm:discrete1} can be regarded as the density of infostation in the highway segment, and the second term is the sum of car densities over all classes except the observer's class. It is interesting to find that the individual throughput depends only on the density of other nodes. Note that the density of nodes belonging to the same class is irrelevant because there will not be any intra-class encounter.

The per-node throughput can also be expressed as
\begin{align}
E[C_i] &= R_p r \left( \frac{1}{d} -\frac{1}{2} \rho_i  + \frac{1}{2}\sum_{m=1}^M \rho_m \right).  \\
&= R_p r \left( \frac{1}{d} -\frac{\lambda p_i }{2 |v_i|} + \frac{1}{2}\sum_{m=1}^M \rho_m \right).
\end{align}

We observe the following:
\begin{itemize}
\item {\em Low-Density Gain:} The class of cars that has the lowest density get the largest
average per-node throughput.

\item {\em High-Speed Gain:} If the speed distribution is equiprobable, i.e., $p_1= p_2=\cdots = p_M$, then the faster the car, the higher average throughput it gets.
\end{itemize}

Now let $C$ be the average per-node throughput. By averaging the per-node throughput in Theorem~\ref{thm:discrete1} over all velocity classes, we have

\begin{align}
E[C] &= \sum_{i=1}^M p_i E[C_i] \\
& = R_p r \Big[ \frac{1}{d} +  \sum_{i=1}^M p_i \Big( \frac{1}{2}
\sum_{m \in \Mo} \rho_m \Big)\Big] \label{eq:main2},
\end{align}
which can be rewritten as follows:

\begin{myth}
\begin{align}
E[C] &= R_p r \left( \frac{1}{d} -\frac{\bar{\rho}}{2} + \frac{1}{2}\sum_{m=1}^M \rho_m \right) \\
&= R_p r\Big[ \frac{1}{d} + \frac{\lambda}{2} \sum_{i \neq j} p_i p_j \Big(
\frac{1}{|v_i|} + \frac{1}{|v_j|} \Big) \Big],
\end{align} \label{thm:discrete2}
where $\bar{\rho} = \sum_i p_i \rho_i$.
\end{myth}

Note that system throughput varies linearly with $C$ and can be obtained by multiplying $C$ with number of users.
Based on the above result, the following facts can be observed:
\begin{itemize}

\item
{\em Incrementally Linear Scalability:} The average per-node throughput increases with the node arrival rate,~$\lambda$, in an incrementally linear fashion.

\item
{\em Mobility Reduces Throughput:} If all cars move faster, then the average per-node throughput decreases.
For example, suppose all cars double their speeds. Then the car density of each velocity class decreases by one half. According to Theorem~\ref{thm:discrete1}, the throughput of all users decreases. Hence the system throughput decreases.

\end{itemize}

Although the velocity of the cars cannot be controlled by the system, it is interesting to know which probability mass function maximizes system throughput, for a given velocity vector  $\{v_1, v_2, \ldots, v_M\}$. We answer this question in the Appendix.

\subsection{Continuous Velocity Distribution}

The analysis for discrete velocity can be extended to the case where the velocity distribution is continuous. This model, called the {\em wide motorway model} in~\cite{kingman}, is applicable to the scenario where there are multiple lanes and moderate traffic. Since nodes can overtake others at different lanes, there is no interaction among the nodes even if they travel in the same direction. A node can have any speed the driver likes, subject to the speed limit.

Suppose that the velocity $V$ is a continuous random variable, whose probability density function
is $f_V(v)$, defined for $v \in [a,b]$.  We divide the interval $[a,b]$ into many intervals, each of length $\Delta v$.
Each interval is approximated by a constant function. We assume that $f_V$ is Lipschitz continuous, so that we can approximate $f_V$ as close as we like by increasing the number of intervals. The next theorem is analogous to Theorem~\ref{thm:discrete1} and~\ref{thm:discrete2}.

\begin{myth} Let $C_i$ denote the throughput of a particular observer node with velocity $v_i$, and $C$ the average per-node throughput. Let $N$ be the number of cars in a highway segment of length~$d$. Then, for all $i$,
\begin{align}
E[C] = E[C_i] &= R_p r\Big(\frac{1}{d}  + \frac{\lambda}{2} E\left[\frac{1}{|V|}\right]  \Big) \label{eq:main3a}\\
&= R_p r\Big(\frac{1}{d} + \frac{1}{2} \frac{E[N]}{d}  \Big). \label{eq:main3b}  \end{align}
\end{myth}

\begin{proof}
As the number of intervals that partition $[a,b]$ approaches infinity, we can rewrite \eqref{eq:discrete1} as

\begin{equation}
E[C_i] = R_p r\Big( \frac{1}{d}  + \frac{\lambda}{2} \int_a^b f_V(v) \frac{1}{|V|} dv \Big),
\end{equation}
which is equal to the right hand side of~\eqref{eq:main3a}. Let $T$ be the random variable $d/V$, which is the duration that a car of velocity $V$ stays in this highway segment.
Conditioned on the velocity, the number of cars of velocity $V$ is Poisson distributed with mean $E[N|V] = \lambda |T|$. Hence
\begin{align}
E[C_i] &= R_p r\Big( \frac{1}{d}  +  \frac{1}{2} \int_a^b f_V(v) \frac{\lambda |T|}{d} dv \Big) \\
&= R_p r\Big( \frac{1}{d}  +   \frac{1}{2} \frac{E[ E[N|V]]}{d}\Big),
\end{align}
which is~\eqref{eq:main3b}. Since $E[C_i]$ is independent of the velocity of the observer, we have $E[C] = E[C_i]$.
\end{proof}

Note that $E[N]/d$ is the car density on the highway.
Based on this theorem, we have the following observations.
The first one is the same as that in the case of discrete speed.
The second one is similar but not exactly the same. The last two observations are different.

\begin{itemize}
\item
{\em Incrementally Linear Scalability:} The average per-node throughput increases with the node arrival rate,~$\lambda$, in an incrementally linear fashion.

\item
{\em Mobility Reduces Throughput:} The average per-node throughput
changes at a rate~$O(E[1/|V|])$. It means that the higher the
mobility, the lower the car density, and the lower the
average per-node throughput.

\item
{\em Perfect Fairness:} $E[C_i]$ is independent of $v_i$. It means
that given the same background traffic on the highway, the throughput
of a node is independent of its own speed. In other words, all
nodes yield the same average throughput, which is different from the
case of discrete speed.

\item
{\em Equivalence of Forward and Reverse Traffics:} The average
throughput of a particular node yielded by encountering with forward traffic is
the same as that yielded by encountering with reverse traffic,
provided that the arrival rates and speed distributions of the two
directions are the same.
\end{itemize}

\section{Concluding Remarks}

We have analyzed the effect of mobility on the performance of a VANET.
Based on the Poisson arrival process, we derive simple formulae for throughput under
both discrete and continuous velocity distribution. There are two major results:
First, system throughput increases linearly with the arrival rate of vehicles. In
other words, the system is linearly scalable. Second, system throughput decreases when
all vehicles increase their speeds, implying that higher overall mobility is not beneficial.

We have also investigated the throughput of individual users. For the discrete velocity case,
the class of users having higher mobility and lower density has higher throughput. In contrast, for the continuous velocity case, all users have the same throughput.

In our analysis, we assume that at most two cars meet each other at
any time instant. If the traffic density is high, this assumption may not
hold. For example, a node can overhear the transmission of
other nodes. However, the performance of the system in such a
scenario depends on the details of a particular transmission
protocol, such as how transmission is initiated and how transmission
conflicts are resolved. This is not within the scope of our
framework and we leave it for future research.

\appendix

\section{Optimal Probability Mass Function for the Case of Discrete Velocity}

Given $\{v_1, v_2, \ldots, v_M\}$, we would like to know what the optimal probability mass function is. The problem can be formally stated as follows.
\begin{align*}
\mbox{Maximize }  & F(\bp) =  \sum_{i \neq j} p_i p_j \left( \frac{1}{|v_i|} + \frac{1}{|v_j|} \right) \\
\mbox{subject to } & \sum_{m=1}^M p_m = 1, \mbox{ and } p_i \geq 0 \;\; \forall i.
\end{align*}

To check whether this is a convex optimization problem, we first eliminate the equality constraint. We consider the function
\begin{equation}
G(p_1, \ldots, p_{M-1}) \equiv F(p_1, \ldots, p_{M-1}, 1 - \sum_{i=1}^{M-1} p_i).
\end{equation}
It can be shown that the Hessian of $G$ is given by
\begin{equation}
-2 \mbox{ diag}(\frac{1}{|v_1|}, \ldots, \frac{1}{|v_{M-1}|}) - \frac{2}{|v_M|} \bJ_{M-1},
\end{equation}
where diag($\bx$) is the diagonal matrix with diagonal elements given by $\bx$, and $\bJ_n$ is the $n\times n$ all-one matrix. Since the first matrix is negative definite and the second one is negative semi-definite, the Hessian of $G$ is negative definite. Hence, the function $G$ is strictly concave, and there is one unique optimal point, $\bp^*$.

\begin{myprop} \label{pr:p_decrease}
If $|v_1| \leq |v_2| \leq \cdots \leq |v_m|$, then at the optimal point $\bp^*$, we must have
\begin{equation}
p_1 \geq p_2 \cdots \geq p_M.
\end{equation}
\end{myprop}

\begin{proof}
Suppose $p^*_i < p^*_j$, where $i < j$. Consider another point $\bp'$, with all components the same as $\bp^*$ except the $i$-th and the $j$-th components swapped. By definition, it can
be seen that $F(\bp^*) < F(\bp')$.
\end{proof}

Now we try to solve the optimization problem. Introduce the Lagrangian multipliers $\lambda_i$'s, $i=1, 2, \ldots, M$, for the non-negative constraints and $\nu$ for the equality constraint. The KKT conditions after eliminating $\lambda_i$'s become
\begin{gather}
p_i (\nu - \sum_{k\neq i} p_k \alpha_{ik}) = 0  \;\; i = 1, \ldots, M, \label{eq:KKT}\\
\sum_{m=1}^M p_m = 1, \mbox{ and } p_i \geq 0 \;\; i = 1, \ldots, M,
\end{gather}
where $\alpha_{ik} = 1/|v_i| + 1/|v_k|$.

Let $\bA$ be the $M\times M$ matrix whose diagonal components are all zero and $(i,j)$-th component equal to $\alpha_{ik}$ for $i\neq k$. Let $\bA_k$ be its leading principal submatrix of order $k$, that is, its last $M-k$ rows and $M-k$ columns are deleted. Let $\one_k$ be the $k \times k$ all-one vector, and $\bp_k$ be the first $k$ components of $\bp$. Without loss of generality, we assume that $|v_1| \leq |v_2| \leq \cdots \leq |v_m|$. The optimal solution can be found by the following algorithm.
\begin{myalg}
(Initialization) Let $n := M$.
\begin{enumerate}
\item Compute
$$
\bp_n = \frac{\bA_n^{-1}\one_n}{\|\bA_n^{-1}\one_n\|_1},
$$
where $\|\cdot \|_1$ is the $l_1$ norm.
\item If all components of $\bp_n$ are non-negative, then output $\bp_M$. Otherwise, let $p_n := 0$ and $n := n - 1$. Repeat step 1.
\end{enumerate}
\end{myalg}

This algorithm produces the correct solution because it simply tries to solve \eqref{eq:KKT}, assuming that $p_i \neq 0$ for all $i$. Note that the multiplier $\nu$ is adjusted such that the equality constraint is satisfied. This corresponds to the normalization factor in the algorithm.
If all the components are non-negative, then the KKT conditions are satisfied. Otherwise, one of the $p_i$'s must be zero because of \eqref{eq:KKT}. By Proposition~\ref{pr:p_decrease}, the last component must be zero. Hence, we reduce the dimension of the problem by one and then repeat.

\begin{table}
\caption{Optimal Probability Mass Functions for Different Velocity Values.} \label{ta:pmf}
\begin{center}
\begin{tabular}{|c|ccccc|} \hline \hline
Speed & 80 & 90 & 100 & 110 & 120 \\
\hline Distribution & 0.26 & 0.23 & 0.2 & 0.17 & 0.14 \\
\hline \hline
Speed & 50 & 60 & 70 & 80 & 130 \\
\hline Distribution & 0.3077 & 0.2692 & 0.2308 & 0.1923 & 0 \\
\hline \hline
Speed & 20 & 30 & 40 & 110 &120 \\
\hline Distribution & 0.3889 & 0.3333 & 0.2778 & 0 & 0 \\
\hline \hline
\end{tabular}
\end{center}
\end{table}

For example, we consider the situation where the nodes can be divided into five classes. Given the velocity values, we can compute the optimal probability mass function. Three examples are shown in Table~\ref{ta:pmf}. It can be seen that some of the $p_i$'s can be equal to zero. However, this occurs only for some extreme cases. We have tested many other cases. Typically, all $p_i$'s will be greater than zero.

Besides, it can be shown that at least two classes must have probabilities strictly greater than zero, for otherwise no encounter in the system can occur. According to Proposition~\ref{pr:p_decrease}, they must be $p_1$ and $p_2$. If there are only two classes or $p_i = 0$ for $i \geq 3$, then it can be easily shown that $p_1^* = p_2^* = 0.5$ is optimal.




\end{document}